\documentclass[preprint2]{aastex}
\usepackage{url}
\usepackage{epsfig}

\begin{document}
\title{\large{\rm{TOWARDS A DETERMINATION OF DEFINITIVE PARAMETERS FOR THE LONG PERIOD CEPHEID S~VULPECULAE}}}

\author{\sc \small D. G. Turner}
\affil{\footnotesize{Department of Astronomy and Physics, Saint Mary's University, Halifax, Nova Scotia B3H 3C3, Canada.}}
\email{turner@ap.smu.ca}

\begin{abstract}
A new compilation of {\it UBV} data for stars near the Cepheid S~Vul incorporates {\it BV} observations from APASS and NOMAD to augment {\it UBV} observations published previously. A reddening analysis yields mean colour excesses and distance moduli for two main groups of stars in the field: the sparse cluster Turner~1 and an anonymous background group of BA stars. The former appears to be $1.07\pm0.12$ kpc distant and reddened by $E_{B-V}=0.45\pm0.05$, with an age of $10^9$ yrs. The previously overlooked latter group is $3.48\pm0.19$ kpc distant and reddened by $E_{B-V}=0.78\pm0.02$, with an age of $1.3\times 10^7$ yrs. Parameters inferred for S~Vul under the assumption that it belongs to the distant group, as also argued by 2MASS data, are all consistent with similar results for other cluster Cepheids and Cepheid-like supergiants.
\end{abstract}

\section{{\rm \footnotesize INTRODUCTION}}

The variability of S~Vulpeculae was first recognized in 1836-37 with an estimated cycle length of $\sim$68 days. The star was at times classified as either semi-regular or RV Tauri-type (e.g., Joy 1952). A study by Nassau \& Ashbrook (1943) appears to be the first to identify the variations as those of a long-period Cepheid. Photoelectric studies by Fernie (1970) and others (Berdnikov 1993, 1994; Heiser 1996) subsequently confirmed the Cepheid nature of S~Vul, making it the longest period classical Cepheid recognized in the Galaxy. A few other Cepheid-like supergiants (e.g., V810~Cen, HD~18391) have periods in excess of 100 days. Like S~Vul, they have small light amplitudes.

\begin{figure}[t]
\centerline{
\epsfig{file=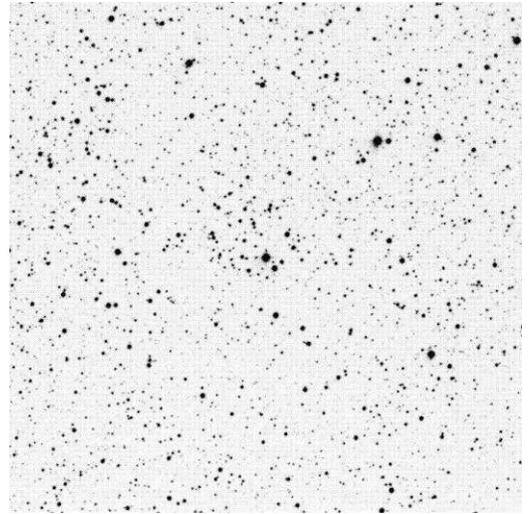, scale=0.60}
}
\caption{\small{The 15$^{\prime}\times15^{\prime}$ field of S~Vul from the Palomar Observatory Sky Survey blue image.}}
\label{fig1}
\end{figure}

In the 1970s the possibility was raised that S~Vul might be a member of the association Vul OB2 (Tsarevskii 1971; Turner 1980), a link that would provide a means of establishing both the luminosity and intrinsic colour of S~Vul from the distance and reddening of association stars. Circa 1980 the author noticed that S~Vul lies in an anonymous open cluster (Fig.~\ref{fig1}), now catalogued as Turner~1 (Turner 1985). A photometric study of the cluster was subsequently performed using photoelectric and photographic photometry (Turner et al.~1986), the latter calibrated using stars in the photoelectric sequence and secondary images offset by $\sim$4$^{\rm m}.6$ from the primary images using a Racine-Pickering wedge on the 3.6-m Canada-France-Hawaii Telescope. Interpreting the photometry was complicated, but it was possible to isolate an old cluster of FG dwarfs $\sim$650 pc distant reddened by $E_{B-V}\simeq0.50$, as well as possible background B stars of larger reddening. The presence of the background group of B dwarfs was later confirmed from 2MASS infrared photometry (Turner 2011), but has not yet been studied using optical photometry. This paper attempts to rectify that omission.

\section{{\rm \footnotesize DATA ANALYSIS}}

\begin{figure}[t]
\centerline{
\epsfig{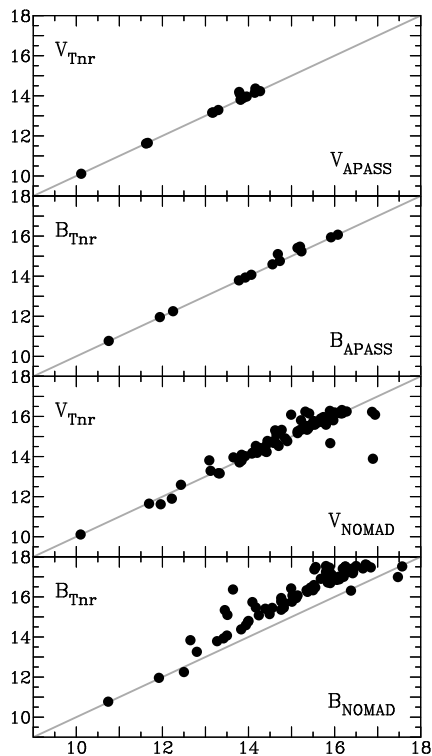}
}
\caption{\small{A comparison of {\it BV} magnitudes from Turner et al. (1986) with data from the APASS (upper) and NOMAD (lower) surveys. Gray lines represent the situation for exact coincidence.}}
\label{fig2}
\end{figure}

Optical studies of reddening in Galactic star fields are best done using Johnson system {\it UBV} photometry, for which interstellar reddening is readily separated from other effects (temperature, metallicity, etc.). New {\it U}-band observations are not yet available for the S~Vul field, but the availability of {\it BVg$^{\;\prime}$r$^{\;\prime}$i$^{\;\prime}$} observations for the field from the AAVSO Photometric All-Sky Survey (APASS) provides a useful beginning. A comparison of previous photoelectric and photographic {\it V}-band and {\it B}-band data for the S~Vul field with APASS data is provided in the upper portion of Fig.~\ref{fig2}. A similar comparison with {\it BV} data from the Naval Observatory Merged Astrometric Dataset (NOMAD, Zacharias et al. 2005) is shown in the lower portion of Fig.~\ref{fig2}. NOMAD data are derived from scans of plates in the Palomar Observatory Sky Survey.

The APASS photometry is in excellent agreement with the Turner et al. (1986) photometry, except for three stars in the APASS survey that may be affected by crowding. There is also a lack of APASS data for many stars near S~Vul within the APASS limits, which may also be a result of image crowding. The next release of APASS for the S~Vul field may overcome such problems.

\begin{figure}[t]
\centerline{
\epsfig{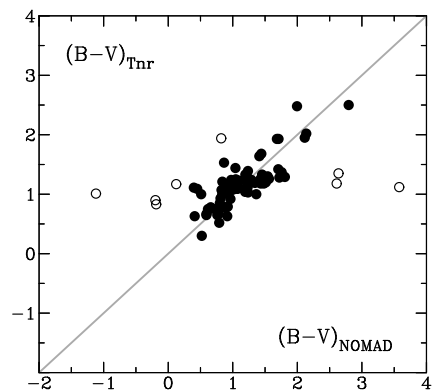}
}
\caption{\small{A comparison of {\it B--V} colours from Turner et al. (1986) with colours resulting from the adjusted NOMAD photometry. A gray line represents exact coincidence, while open circles indicate discrepant points not used for the best fit.}}
\label{fig3}
\end{figure}

The comparison of previous photoelectric and photographic {\it V}-band and {\it B}-band data for the S~Vul field with NOMAD data reveals sizable systematic offsets, particularly in {\it B}. Such offsets appear occasionally in other NOMAD fields and may be linked to the choice of calibration stars. It is possible to infer from the plots the main trend in the data in order to correct the NOMAD results for S~Vul stars for the calibration error. But correcting the resulting NOMAD {\it B--V} colours to the photoelectric system is rather complicated. Fig.~\ref{fig3} is a comparison of the adjusted NOMAD {\it B--V} colours with the photoelectric and photographic colours in the survey of Turner et al. (1986). The adopted trend relation (not shown) depends significantly upon how many of the most discrepant stars are omitted from the fit.

\begin{figure}[t]
\centerline{
\epsfig{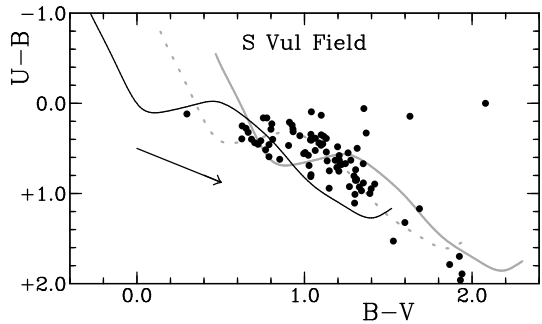}
}
\caption{\small{Colour-colour diagram for stars near S~Vul with full {\it UBV} data. The intrinsic relation for dwarfs is the black polynomial, while the dotted gray relation and the solid gray relation represent the intrinsic relation reddened by $E_{B-V}=0.45$ and $E_{B-V}=0.78$, respectively. An arrow denotes the reddening relation for the field (Turner 1980, 1989).}}
\label{fig4}
\end{figure}

Analysis of the combined photoelectric and photographic UBV, APASS BV CCD, and NOMAD BV observations is involved, but a solution was found. Data from the various sources were assigned weights depending upon the expected precision, with highest weights for the photoelectric observations, medium weights for the CCD and photographic {\it BV} observations, and low weights for the NOMAD {\it BV} data. The {\it U}-band data from the Turner et al.~(1986) study were adopted as given. In other words, the APASS and NOMAD data were used to improve the {\it V}-magnitudes and {\it B--V} colours for stars in the Turner et al. (1986) study, as well as to add {\it BV} data for faint stars lying beyond its spatial and magnitude limits. The {\it UBV} colours for the resulting sample of stars are plotted in the colour-colur diagram of Fig.~\ref{fig4}.

The reddening relation for the field of S~Vul is well established from previous studies (Turner 1980, 1989) to be described by the relation $E_{U-B}/E_{B-V} = 0.74 + 0.02\;E_{B-V}$, which has been approximated by $E_{U-B}/E_{B-V} \simeq 0.75$ in Fig.~\ref{fig4}. As found previously (Turner et al.~1986), most stars in Turner~1 are FG dwarfs reddened by $E_{B-V}\simeq0.48$. A best fit by eye to the data of Fig.~\ref{fig4} yields a mean reddening of $E_{B-V}=0.45\pm0.05$ for the FG dwarfs, where the uncertainty represents scatter arising from differential reddening in the field (Turner 1980), uncertainties in the observations, and unknown corrections for ultraviolet excesses in the G dwarfs resulting from line blanketing (Wildey et al. 1962). There is also a relatively clear sequence of B dwarfs reddened by $E_{B-V}=0.78\pm0.02$ that represents background stars detected beyond the foreground cluster. Since the stars are dispersed across the field, the scatter in their reddenings may also arise from the small amount of differential reddening in the field.

\begin{figure}[t]
\centerline{
\epsfig{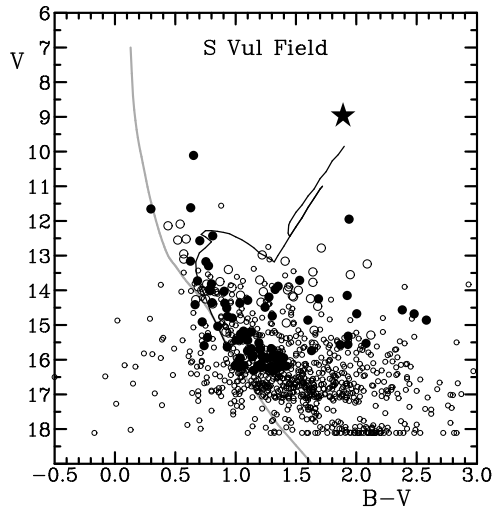}
}
\caption{\small{Colour-magnitude diagram for Turner~1 with $E_{B-V}=0.45$ and stars from the Turner et al. (1986) survey (filled circles), APASS (open circles), and NOMAD (small open circles). S~Vul is plotted as a star symbol, the gray curve is the ZAMS for $V-M_V=11.50$, and a black curve is an isochrone for $\log t= 9.0$.}}
\label{fig5}
\end{figure}

The colour-magnitude diagram for the stars in Turner~1 can be interpreted with the established reddenings of Fig.~\ref{fig4}. Fig.~\ref{fig5} presents the colour-magnitude diagram for the analyzed stars using a zero-age main sequence (ZAMS) for the smaller reddening of $E_{B-V}=0.45$. The ZAMS matches the lower envelope of FG stars at $V-M_V\simeq11.50\pm0.20$. With the adopted reddening, the intrinsic distance modulus is $V_0-M_V=10.15\pm0.25$, corresponding to a distance of $1.07\pm0.12$ kpc. That is larger than the value of 643 pc found in the study by Turner et al.~(1986), the difference being accounted for by the fainter stars added here to the ZAMS for the cluster. An age isochrone for $\log t=9.0$ also provides a good fit to the evolved portion of the main sequence, as well as to a putative group of G giant members in Fig.~\ref{fig5}. The eye fit to the data is limited by the mixed quality of the photometry. A deeper CCD study is in progress to confirm the present results.

\begin{table}[t]
\caption{Intrinsic Parameters for S~Vulpeculae.}
\begin{tabular}{lcc}
\hline \hline
Parameter &Value &Reference \\
\hline
Period, {\it P} &$68^{\rm d}.0194$ &Berdnikov (1994) \\
$\log P$ &1.8326 &Berdnikov (1994) \\
$\langle V \rangle$ &8.974 &Berdnikov (2007) \\
$\langle B \rangle-\langle V \rangle$ &1.890 &Berdnikov (2007) \\
$E_{B-V}$(B0) &0.78 &This Paper \\
$E_{B-V}$(C$\delta$) &0.695 &This Paper \\
$(\langle B\rangle$--$\langle V\rangle)_0$ &1.195 &This Paper \\
$\log T_{\rm eff}$ &3.651 &This Paper \\
Distance, {\it d} &$3.48 \pm0.19$ kpc &This Paper \\
$\log t$ &7.1 &This Paper \\
$\langle M_V \rangle$ &$-6.08 \pm0.12$ &This Paper \\
$\langle M_{\rm bol}\rangle$ &$-6.40 \pm0.12$ &This Paper \\
$\langle M[3.6] \rangle$ &$-8.51 \pm0.12$ &This Paper \\
$\langle M[4.5] \rangle$ &$-8.42 \pm0.12$ &This Paper \\
\hline
\end{tabular}
\label{tab1}
\end{table}

\begin{figure}[t]
\centerline{
\epsfig{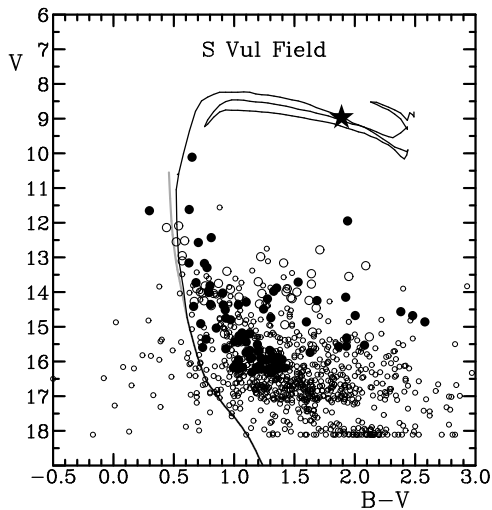}
}
\caption{\small{A repeat of Fig.~\ref{fig5} for the same data, with the ZAMS (gray relation) plotted using $E_{B-V}=0.78$ and $V-M_V\simeq15.05$. An isochrone for $\log t=7.1$ is plotted as a black curve.}}
\label{fig6}
\end{figure}

A colour-magnitude diagram for the same stars is presented in Fig.~\ref{fig6} for the established reddening of the B stars, namely $E_{B-V}=0.78$. Here a reasonable match of the ZAMS to faint B and A stars can be made for $V-M_V\simeq15.05\pm0.10$. With the adopted reddening, the intrinsic distance modulus is $V_0-M_V=12.71\pm0.12$, corresponding to a distance of $3.48\pm0.19$ kpc. That is less than the distance of 4.41 kpc found by Turner~(1980) for Vul OB2, and the stars are also somewhat older than is the case for the luminous OB stars populating Vul OB2. An age isochrone for $\log t=7.1$ provides a good fit to the evolved portion of the main sequence, as well as for S~Vul. Again, a deeper CCD study is in progress to confirm the results.

\begin{table}[t]
\caption{Properties of Clusters Containing Pulsating Supergiants.}
\begin{tabular}{@{\extracolsep{-1mm}}llccc}
\hline \hline
Cluster &Cepheid &$\log P$ &$\log t$ &$M_{\rm RTO}$ \\
\hline
Alessi~95 &SU~Cas &0.443 &8.20 &$4.1 M_{\odot}$ \\
Harrington~1 &$\alpha$~UMi &0.599 &7.90 &$5.4 M_{\odot}$ \\
Platais~1 &V1726~Cyg &0.627 &8.00 &$4.9 M_{\odot}$ \\
Berkeley~58 &CG~Cas &0.640 &8.00 &$4.9 M_{\odot}$ \\
NGC~7790 &CEb~Cas &0.651 &8.00 &$4.9 M_{\odot}$ \\
NGC~1647 &SZ~Tau &0.652 &8.10 &$4.2 M_{\odot}$ \\
NGC~7790 &CF~Cas &0.688 &8.00 &$4.9 M_{\odot}$ \\
NGC~7790 &CEa~Cas &0.711 &8.00 &$4.9 M_{\odot}$ \\
van den Bergh~1 &CV~Mon &0.731 &7.90 &$5.4 M_{\odot}$ \\
NGC~6067 &QZ~Nor &0.733 &7.80 &$5.9 M_{\odot}$ \\
NGC~6649 &V367~Sct &0.799 &7.85 &$5.7 M_{\odot}$ \\
Collinder~394 &BB~Sgr &0.822 &7.85 &$5.7 M_{\odot}$ \\
NGC~129 &DL~Cas &0.903 &7.85 &$5.7 M_{\odot}$ \\
NGC~6087 &S~Nor &0.989 &7.80 &$5.9 M_{\odot}$ \\
Lyng\aa~6 &TW~Nor &1.033 &7.80 &$5.9 M_{\odot}$ \\
NGC~6067 &V340~Nor &1.053 &7.80 &$5.9 M_{\odot}$ \\
Teutsch~106 &GT~Car &1.119 &7.70 &$6.5 M_{\odot}$ \\
Trumpler~35 &RU~Sct &1.295 &7.55 &$7.9 M_{\odot}$ \\
Turner~2 &WZ~Sgr &1.339 &7.55 &$7.9 M_{\odot}$ \\
Anon~Vel~OB &SW~Vel &1.370 &7.50 &$8.3 M_{\odot}$ \\
Turner~14 &AQ~Pup &1.477 &7.50 &$8.3 M_{\odot}$ \\
Turner~8 &SV~Vul &1.653 &7.20 &$12.2 M_{\odot}$ \\
Turner~1b &S~Vul &1.837 &7.10 &$14.2 M_{\odot}$ \\
Stock~14 &V810~Cen &2.184 &6.90 &$20.4 M_{\odot}$ \\
Anon &HD~18391 &2.250 &6.90 &$20.4 M_{\odot}$ \\
Berkeley~87 &BC~Cyg &2.841 &6.70 &$31.3 M_{\odot}$ \\
Trumpler~37 &$\mu$~Cep &3.002 &6.60 &$39.7 M_{\odot}$ \\ 
\hline
\end{tabular}
\label{tab2}
\end{table}

The sequence of faint, reddened B dwarfs displays no degree of concentration towards the centre of Turner~1, so must represent a sparse clumping of young stars along the line of sight to the Cepheid. Similar conclusions were reached in a study of the field that made use of 2MASS observations (Turner 2011). The average reddening of $E_{B-V}=0.78\pm0.02$ for these stars must also apply to the Cepheid, in which case its stellar reddening becomes $E_{B-V}=0.695$ according to the relationship of Fernie (1970). If S~Vul is a member of the same group, as seems likely, it has an inferred luminosity of $\langle M_V\rangle=-6.08\pm0.12$. Other properties of S~Vul as a likely member of the background group of B stars are summarized in Table~\ref{tab1}.

\begin{figure}[t]
\centerline{
\epsfig{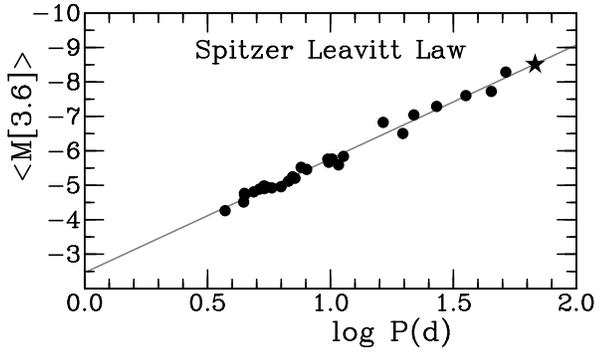}
}
\caption{\small{Absolute Spitzer $3.6\mu$m magnitudes for parallax and cluster Cepheids tied to field reddenings for the calibrators. S~Vul is depicted by a star symbol.}}
\label{fig7}
\end{figure}

The consistency of the derived luminosity of S~Vul with those of other calibrating Cepheids is seen in its absolute Spitzer $3.6\mu$m magnitude ([3.6]) in Fig.~\ref{fig7}, where the observations for cluster and parallax Cepheids are taken from Monson et al. (2012), but modified to the system of field reddenings for the calibrators. Similar results apply to the Spitzer $4.5\mu$m magnitude. 

Similar conclusions are reached if one compares the implied mass and age of S~Vul with empirical values derived for cluster Cepheids. A preliminary study of such characteristics by Turner (1996) has been updated in Table~\ref{tab2} using new results for Cepheids and Cepheid-like supergiants associated with open clusters (Turner 2010; Turner et al. 2009, 2012a, 2012b; Majaess et al. 2013). Included are results for two red supergiant variables associated with open clusters: BC~Cyg in Berkeley~87 and $\mu$ Cep in Trumpler~37.

\begin{figure}[t]
\centerline{
\epsfig{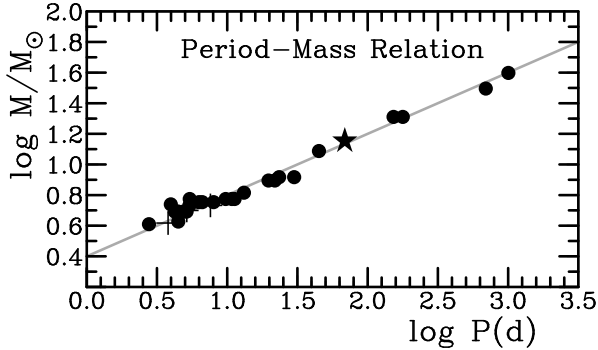}
}
\caption{\small{Evolutionary (points) and dynamical (plus signs) masses for Cepheids and Cepheid-like supergiants. S~Vul is plotted as a star symbol, and the linear relation is from Turner (2012).}}
\label{fig8}
\end{figure}

Dynamical masses have been derived for three Cepheids in binary systems (Evans et al. 2009, 2011; Pietrzy\~{n}ski et al. 2010), while evolutionary masses have been deduced for Cepheids and Cepheid-like objects in clusters using $M_{RTO}$, the mass of stars at the red edge of the tip of the evolved main sequence, i.e., stars in the terminal stages of core hydrogen burning prior to the first crossing of the instability strip, as inferred from the models of Meynet et al.~(1993). In the study by Turner (1996) the evolutionary masses of cluster Cepheids were assumed to be roughly 10\% larger than values of $M_{RTO}$, but the close coincidence in Fig.~\ref{fig8} of $M_{RTO}$ evolutionary masses with dynamical masses for binary Cepheids suggests that the two estimates may be almost identical (see also Turner 2012). The evolutionary stages of shell hydrogen burning and core helium burning therefore occur on significantly shorter time scales than is the case for core hydrogen burning in stars. The implied evolutionary mass of S~Vul from its association with BA stars in its vicinity closely agrees with evolutionary masses deduced for other cluster Cepheids (Fig.~\ref{fig8}).

\begin{figure}[t]
\centerline{
\epsfig{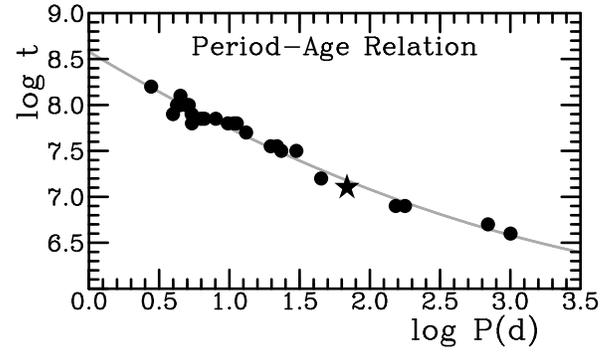}
}
\caption{\small{Empirical period-age relationship for Cepheids and Cepheid-like supergiants in open clusters (points), including S~Vul (star symbol). The gray polynomial is the best-fitting relationship to the data.}}
\label{fig9}
\end{figure}

A similar result is found for the empirical Cepheid period-age relation, as deduced for cluster Cepheids and pulsating red supergiants in Table~\ref{tab2} and plotted in Fig.~\ref{fig9}. The pulsation periods used in Table~\ref{tab2} and plotted in Figs.~\ref{fig7},~\ref{fig8}, and~\ref{fig9} are those for fundamental mode pulsation. The review by Turner (2012) and other studies of the Cepheid period-age correlation assume that the relationship is linear, but the results of Fig.~\ref{fig9} suggest that the relationship may be non-linear for high-mass pulsators. Such non-linearity depends upon how closely red supergiant pulsators resemble normal Cepheids.

\section{{\rm \footnotesize SUMMARY}}
Fundamental properties have been deduced for S~Vul from its likely association with a group of B stars in its immediate vicinity. All derived parameters for the Cepheid are consistent with its membership in the loose group of B stars, but confirmation must await a deeper {\it UBV} study of stars surrounding the Cepheid and measurement of radial velocities for brighter association members. Preliminary measurements have already been made of potential candidates using the Plaskett telescope at the Dominion Astrophysical Observatory: older measurements from the 1980s by Doug Welch (unpublished), and more recent CCD observations. Results will be presented elsewhere. The location of S~Vul at the extreme long-period end of the Leavitt relation for classical Cepheids makes it extremely valuable for anchoring the slope of the relationship, so further study is essential.

The present study expands upon results described in an oral paper given at the spring meeting of the AAVSO held in Boone, North Carolina, May 17--18, 2013.

\section*{{\rm \footnotesize REFERENCES}}
\small{
Berdnikov, L.N.: 1993, {\it AstL}, {\bf 19}, 84.\\
Berdnikov, L.N.: 1994, {\it AstL}, {\bf 20}, 232.\\
Berdnikov, L.N.: 2007, \url{http://www.sai.msu.ru/groups/cluster/CEP/PHE} \\
Evans, N.R., Massa, D., Proffitt, C.: 2009, {\it AJ}, {\bf 137},
 3700.\\
Evans, N.R., et al.: 2011, {\it AJ}, {\bf 142}, 87.\\
Fernie, J.D.: 1970, {\it AJ}, {\bf 75}, 244.\\
Heiser, A.H.: 1996, {\it PASP}, {\bf 108}, 603. \\
Joy, A.H.: 1952, {\it ApJ}, {\bf 115}, 25.\\
Mahmoud, F., Szabados, L.: 1980, {\it IBVS}, {\bf 1895}, 1.\\
Majaess, D., et al.: 2013, {\it Ap\&SS}, in press.\\
Meynet, G., Mermilliod, J.-C., Maeder, A.: 1993,
{\it A\&AS}, {\bf 98}, 477.\\
Monson, A.J., et al.: 2012, {\it ApJ}, {\bf 759}, 146. \\
Nassau, J.J., Ashbrook, J.: 1943, {\it AJ}, {\bf 50}, 97.\\
Pietrzy\~{n}ski, G., et al.: 2010, {\it Nature}, {\bf 468}, 542.\\
Turner, D.G.: 1976, {\it AJ}, {\bf 81}, 97.\\
Turner, D.G.: 1980, {\it ApJ}, {\bf 235}, 146.\\
Turner, D.G.: 1985, {\it JRASC}, {\bf 79}, 175.\\
Turner, D.G.: 1989, {\it AJ}, {\bf 98}, 2300.\\
Turner, D.G.: 1996, {\it JRASC}, {\bf 90}, 82.\\
Turner, D.G.: 2010, {\it Ap\&SS}, {\bf 326}, 219.\\
Turner, D.G.: 2011, {\it RMxAA}, {\bf 47}, 127.\\
Turner, D.G.: 2012, {\it JAAVSO}, {\bf 40}, 502.\\
Turner, D.G., Leonard, P.J.T., Madore, B.F.: 1986, {\it JRASC}, {\bf 80}, 166.\\
Turner, D.G., et al.: 2009, {\it AN}, {\bf 330}, 807.\\
Turner, D.G., et al.: 2012a, {\it MNRAS}, {\bf 422}, 2501.\\
Turner, D.G., et al.: 2012b, {\it AJ}, {\bf 144}, 187.\\
Tsarevskii, G.S.: 1971, {\it SvA}, {\bf 15}, 169.\\
Wildey, R.L., et al.: 1962, {\it ApJ}, {\bf 135}, 94.\\
Zacharias N., et al.: 2005, {\it BAAS}, {\bf 36}, 1418.\\}
\end{document}